\documentclass[aps,prb,twocolumn,superscriptaddress,amsmath,amsfonts,email,
               floatfix]{revtex4}

\usepackage{epsfig,psfrag}
\usepackage{CJK}

\usepackage{graphicx}
\usepackage{dcolumn}
\usepackage{bm}
\usepackage{amssymb}

\begin{document}

\title{Energy resonance transfer between quantum defects in metal halide perovskites}
\author{Yu Cui}
\author{Xiao-Yi Liu}
\author{Shi-Yuan Ji}
\author{Yong Sun}
\author{Jia-Pei Deng}
\author{Xu-Fei Ma}
\author{Zhi-Qing Li}
\author{Zi-Wu Wang*}
\affiliation{Tianjin Key Laboratory of Low Dimensional Materials Physics and Preparing Technology, Department of Physics, School of Science, Tianjin University, Tianjin 300354, China}
\email{wangziwu@tju.edu.cn}

\begin{abstract}
Quantum defects have shown to play an essential role for the non-radiative recombination in metal halide perovskites (MHPs). Nonetheless, the processes of charge transfer-assisted by defects are still ambiguous. Herein, we theoretically study the non-radiative multiphonon processes among different types of quantum defects in MHPs using Markvart's model for the induced mechanisms of electron-electron and electron-phonon interactions, respectively. We find that charge carrier can transfer between the neighboring levels of the same type shallow defects by multiphonon processes, but it will be distinctly suppressed with the increasing of the defect depth. For the non-radiation multiphonon transitions between donor- and acceptor-like defects, the processes are very fast and independence of the defect depth, which provide a possible explanation for the blinking phenomena of photoluminescence spectra in recent experiment. We also discuss the temperature dependence of these multiphonon processes and find that their variational trends depend on the comparison of Huang-Rhys factor with the emitted phonon number. These theoretical results fill some gaps of defect-assisted non-radiative processes in the perovskites materials.
\end{abstract}

\maketitle
\section{Introduction}
Metal halide perovskites (MHPs) and their low-dimensional structures are triggering widespread attention
owing to a range of fascinating properties, such as high quantum yield, remarkable photovoltaic conversion
efficiency, high optical absorption coefficient, and ease of fabrication\cite{wx1,wx2,wx3,wx4,wx5}. These properties make MHPs candidates for the next generation of optoelectronic and photovoltaic
devices. However, the inevitable non-radiative recombination processes, dominating the charge carriers
loss, play the crucial roles to determine their basic properties and device performances\cite{wx6,wx7,wx8,wx9,wx10}. There, for example, are trap states which strongly interact with the photogenerated charge carriers, making the conversion efficiencies of perovskite solar cells have not yet reached the thermodynamic limits\cite{wx11,wx12}.
\begin{figure}[htbp]
	\includegraphics[width=3.5in,keepaspectratio]{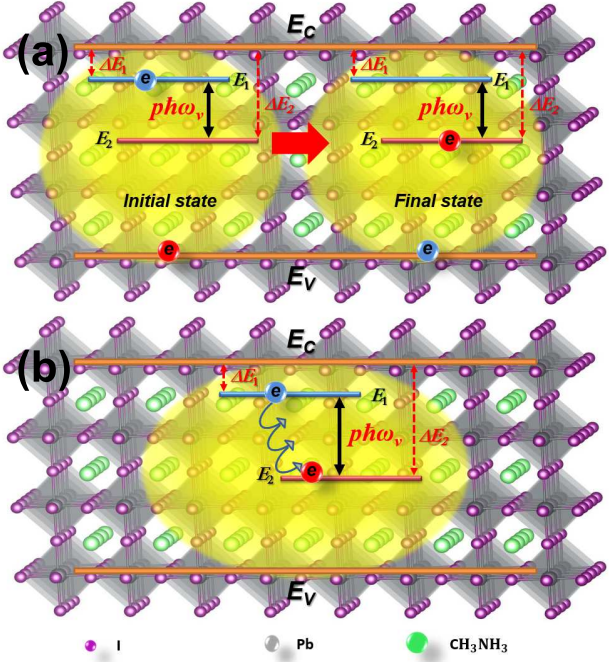}
	\caption{\label{compare} The schematic diagrams of the non-radiative multiphonon processes of energy resonance transfer between quantum defects in CH$_3$NH$_3$PbI$_3$ MHPs. (a) represents a quantum transition induced by electron-electron interaction from the initial state to the final state with its transition energy to be compensated by multiphonon emission $p\hbar {\omega _{\nu }}$, where the initial (final) state consists of two electrons located in the defect level $E_1$ ($E_2$) and in the valence band edge $E_V$, respectively. $\Delta E_1=|E_C-E_1|$ is the initial defect depth, $\Delta E_2=|E_C-E_2|$ is the final defect depth, and $E_C$ denotes the conduct band edge. (b) represents the electron transfer induced by electron-phonon interaction, where the electron relaxes from the initial defect depth $\Delta E_1$ to deeper defect depth $\Delta E_2$ directly by the successive emission of phonons $p\hbar {\omega _{\nu }}$ and $\hbar {\omega _{\nu }}$ denotes the phonon energy.}
\end{figure}
Without a doubt, an essential condition for perovskites-based devices to reach higher performances is the elimination of all loss channels, so the studies of non-radiative loss channels in MHPs are the key topics.

A most concerned non-radiative loss channel is defect-assisted recombination. In the past years, there have been several correlative proofs about defect-assisted processes. de Quilettes $el$ $al.$ have observed the defect states quenched the crystalline fluorescence of discontinuous perovskite films by measuring the microscopic photoluminescence\cite{b1}. From time-resolved photoluminescence spectroscopy, Yamada $et$ $al.$ concluded that the defect-assisted recombination includes non-radiative processes in MHPs\cite{wx defect6}. Bolink $et$ $al.$ also found that the actual devices are hampered by the defect-assisted non-radiative processes according to the voltage dependence of the electroluminescence efficiency and the ideality factor measurements\cite{b2}. In fact, defects (or impurities), especially for the deeper defect levels serving as the trapping centers, play the key role in determining the non-radiative recombination, because of the requirement for higher thermal energy to leave from the traps\cite{b5,b6,b7}, in a way that is ultimately detrimental to the charge transport and opto-electronic properties\cite{b8,b9,b10,b11}. Therefore, the comprehensive understanding of the quantum processes assisted by defects are of crucial for the fundamental properties of MHPs and their potential applications. However, the detail studies on the charge carriers of non-radiative transitions among defect states in these perovskite materials are very few until now.

In the present paper, we study the non-radiative multiphonon processes of energy resonance transfer (ERT) between quantum defects based on Markvart's model\cite{wx moxing1,wx moxing2,wx moxing3,wx moxing4}. The processes between the same types of defects, including donor-donor, acceptor-acceptor and neutral-neutral, as well as between different types of defects, i.e., donor-acceptor pairs, are taken into account. We present the influence of different defect depths on the lifetime of these energy transfer processes induced by the mechanisms of the electron-electron and electron-phonon interactions, respectively, and give the comparisons between them. We also discuss the temperature dependences of these multiphonon transitions among different defects.

\section{theoretical model}
 The processes of the non-radiative multiphonon ERT among quantum defects for the induced mechanisms of electron-electron and electron-phonon interactions are schemed in Fig. 1(a) and (b), respectively. According to the approximation derived by Markvart\cite{wx moxing1,wx moxing2,wx moxing3,wx moxing4}, the transition rate of multiphonon ERT processes $\tau^{-1}$ can be expressed as
\begin{eqnarray}
{\tau_{e-e} ^{{\rm{ - 1}}}}&=&\frac{{{{\left| {{\mathcal{H}_{ij}^{e-e}}} \right|}^2}\sqrt {2\pi } }}{{\hbar \left( {p\hbar {\omega _{LO}}} \right)\sqrt {p\sqrt {1 + {\chi^2_{\lambda}}} } }}\exp [p(\frac{{\hbar {\omega _{LO}}}}{{2k_{B}T}} + \sqrt {1 + {\chi^2_{\lambda}}}\nonumber\\
&&- \chi_{\lambda}\cosh (\frac{{\hbar {\omega _{LO}}}}{{2k_{B}T}}) - \ln (\frac{{1 + \sqrt {1 + {\chi^2_{\lambda}}} }}{\chi_{\lambda}}))],
\end{eqnarray}
with
\begin{eqnarray}
{\left| {{\cal H}_{ij}^{e - e}} \right|^2} = {\left| {\left\langle {{\phi _0}|\langle{\varphi _i}\left| {{H_{e - e}}} \right|{\varphi _j}\rangle|{\phi _0}} \right\rangle } \right|^2},
\end{eqnarray}
\begin{eqnarray}
{H_{e - e}} = \frac{{{e^2}}}{{4\pi {\varepsilon _0}{\varepsilon _r}\left| {{{\bf{r}}_{\bf{1}}}{\bf{ - }}{{\bf{r}}_{\bf{2}}}} \right|}},
\end{eqnarray}
for the electron-electron interaction and
\begin{eqnarray}
{\tau_{e-p} ^{{\rm{ - 1}}}}&=&\frac{{{{\left| {{\mathcal{H}_{ij}^{e-p}}} \right|}^2}\sqrt {2\pi } }}{{\hbar \left( {p\hbar {\omega _{LO}}} \right)\sqrt {p\sqrt {1 + {\chi^2_{\lambda}}} } }}\exp [p(\frac{{\hbar {\omega _{LO}}}}{{2k_{B}T}} + \sqrt {1 + {\chi^2_{\lambda}}}\nonumber\\
&&- \chi_{\lambda}\cosh (\frac{{\hbar {\omega _{LO}}}}{{2k_{B}T}}) - \ln (\frac{{1 + \sqrt {1 + {\chi^2_{\lambda}}} }}{\chi_{\lambda}}))],
\end{eqnarray}
with
\begin{eqnarray}
{{\left| {{\mathcal{H}_{ij}^{e-p}}} \right|}^2}={\left| {\left\langle {\varphi _i\left| {{H_{e - p}}} \right|{\varphi _j}} \right\rangle } \right|^2},
\end{eqnarray}
\begin{eqnarray}
{H_{e - p}} = \sum\limits_{\bf{q}} {{\cal M}(q)(a_{ - {\bf{q}}}^{\dagger} + {a_{\bf{q}}}){e^{i{\bf{q}} \cdot {\bf{r}}}}},
\end{eqnarray}
\begin{eqnarray}
{\cal M} (q)= - i\frac{{\hbar {\omega _{LO}}}}{q}{(\frac{{4\pi \alpha }}{V})^{\frac{1}{2}}}{(\frac{\hbar }{{2{m^*}{\omega _{LO}}}})^{\frac{1}{4}}},
\end{eqnarray}
for the electron-phonon interaction.

\begin{figure*}[htbp]
	\centering
	\includegraphics[width=7in,keepaspectratio]{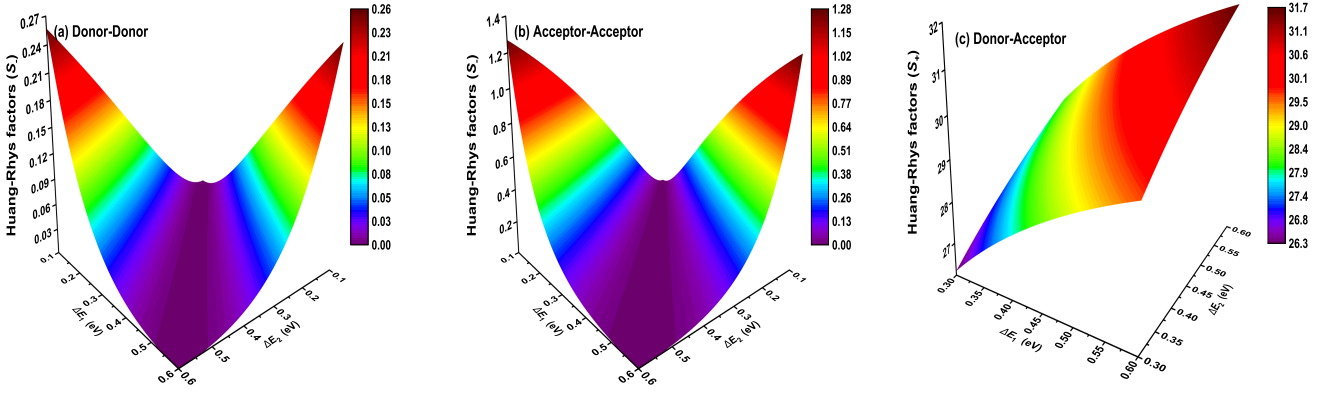}
	\caption{\label{compare} Huang-Rhys factors as functions of the initial defect depths ($\Delta E_1$) and the final defect depths ($\Delta E_2$). (a), (b), and (c) present the donor-donor, acceptor-acceptor, and donor-acceptor processes, respectively.}
\end{figure*}
\newcommand{\tabincell}[2]{\begin{tabular}{@{}#1@{}}#2\end{tabular}}
\begin{table}[htbp]
	\caption{\label{compare} The adopted parameters for the theoretical calculation in MAPbI$_3$.
	}
	\begin{tabular}{ccccccccc}
		\hline
		\hline
		Parameter                  & Symbol  &  Value \\[0.3ex]\hline
		Effective Bohr radius     &   $a^*$  &   8.9 nm\cite{wx canshu1,wx moxing1}   \\
		Effective mass     &   $m^*$  &   0.2\cite{wx canshu1}  \\
		Phonon energy      & $\hbar {\omega _{LO}}$ & 16.5 meV\cite{wx canshu1,wx moxing1} \\
  Fr\"{o}hlich coupling constant &$\alpha$ & 1.73\cite{wx canshu1,wx moxing1} \\		

		\tabincell{c}{Radius of sphere with Brillouin \\ zone volume}  & $q_c$ & 6.2 /nm\cite{wx canshu1}\\
		Lattice constant         & $a_0$ &0.63 nm\cite{wx canshu2}\\
		Vacuum permittivity &${\varepsilon _0}$& 8.85 $\times {10^{ - 12}}$ F/m\cite{wx canshu1}\\
        Free electron (hole) mass &${m_0}$& $9.1 \times {10^{ - 31}}$ kg\cite{wx canshu1}\\
        Band gap &$\left| {{E_C} - {E_V}} \right|$& 1.6 eV\cite{wx moxing1}\\
        Quantum defect parameter         & $\vartheta $ & Variable \\
		Depth of defect              &$\Delta E$  &Variable\\
		Relative permittivity  &${\varepsilon _r}$& 20${\varepsilon _0}$\\
       \hline
	\end{tabular}
\end{table}

Longitudinal optical (LO) phonon is mainly taken into account because of the strong charge carrier-LO phonon coupling in MHPs\cite{wx ep1,wx ep2}. $p$ is the emitted phonon number, $\hbar {\omega _{LO}}$ is the LO phonon energy, $T$ is the temperature and $k_B$ is the Boltzmann constant. $\varphi_i$ ($\varphi_j$) is the initial (final) ground-state wave function of defect with the depth $\Delta E_1$ ($\Delta E_2$), and ${\phi _0}$ is the free carrier state at the band edge. In Eq. (3), ${\varepsilon _0}$ is the permittivity of vacuum, ${\varepsilon _r}$ is the relative permittivity, and $\bf r_1$ as well as $\bf r_2$ are position
variables for electrons. Eq. (6) describes the interaction between charge carrier and LO phonons in the Fr\"{o}hlich mechanism\cite{wx moxing5} with the coupling element $\mathcal{M}(q)$, where ${a_ {\bf q}}$ ($a_{ - {\bf q}}^\dag $) is annihilation (creation) a LO phonon with wave vector $\bf q$, $V$ is the volume, $m^*$ is the effective mass of electron (or hole), and $\alpha$ denotes the Fr\"{o}hlich coupling strength in this  mechanism. The parameter ${\chi_\lambda }$ is defined as
\begin{equation}
{\chi_{\lambda} } = \left\{ \begin{gathered}
\frac{\mathcal{S}_{\lambda}}{{p\sinh (\hbar {\omega _{LO}}/2k_{B}T)}}\quad \hfill  (\mathcal{S}_{\lambda}<p)\\
\frac{p}{{\mathcal{S}_{\lambda}\sinh (\hbar {\omega _{LO}}/2k_{B}T)}}\quad \hfill  (\mathcal{S}_{\lambda}>p)
\end{gathered} \right.,
\end{equation}
where $\mathcal{S}_{\lambda}$ is Huang-Rhys (HR) factor with the subscript $\lambda$ = - ($\lambda$ = +) represents the ERT between the same (different) types of defects, which denotes the average LO phonon numbers around the trapped carrier, reflecting the strength of the electron-phonon coupling\cite{wx Huang3,wx Huang2,wx Huang1}.

Following the quantum defect model\cite{wx q1,wx q2,wx q3,wx q4}, the ground-state wave function $\varphi ({r})$ for defect with arbitrary binding energy can be expressed as
\begin{equation}
\varphi (r)=\mathcal{N}(\frac{r}{a^*})^{\mu-1}\exp({-\frac{r}{\vartheta a^*}}),
\end{equation}
with
\begin{equation}
\mathcal{N}=\frac{1}{\sqrt{4\pi a^{*3}(\vartheta/2)^{2\mu+1}\Gamma(2\mu+1)}},
\end{equation}
\begin{equation}
a^*=\frac{4\pi\varepsilon\hbar^2}{e^2m^*},
\end{equation}
\begin{equation}
\vartheta  = \frac{{{e}}}{{\sqrt {8\pi \varepsilon{a^*}\Delta E} }},
\end{equation}
where $\mathcal{N}$, $a^*$, and $\vartheta$ represent normalization constant, effective Bohr radius, and quantum defect parameter, respectively. $\Delta E$ denotes the defect depth, the type of defect is reflected by the parameter $\mu$: $\mu$ = -$\vartheta$ for the donor-like defect, $\mu$ = +$\vartheta$ for the acceptor-like defect, and $\mu$ = 0 for the neutral defect. Substituting Eqs. (3), (6), (7) and (9) into Eqs. (2) and (5), the matrix elements become
\begin{eqnarray}
{\left| {{\mathcal{H}_{ij}^{e-e}}} \right|^2} &=& \frac{{64{e^4}\mathcal{N}_1^2\mathcal{N}_2^2}}{{\varepsilon _0^2\varepsilon _r^2a_0^6}}|\int {\int {\int {d{r_1}} } } d{r_2}r_1^2r_2^2 \nonumber\\
&& \times \frac{{\sin (k{r_1})\sin (k{r_2})}}{{{r_1}{r_2}{k^2}}}{(\frac{{{r_1}{r_2}}}{{{a^{*2}}}})^{{\mu _1} + {\mu _2} - 2}}\nonumber\\
&& \times \exp ( - \frac{{{r_1}}}{{{\vartheta _1}{a^*}}} - \frac{{{r_2}}}{{{\vartheta _2}{a^*}}}){dk}|^2,
\end{eqnarray}
for electron-electron interaction\cite{wx Huang2,wx ee1,wx ee2,wx ee3}, and
\begin{eqnarray}
|H_{ij}^{e - p}{|^2}{\rm{ }} &=& \frac{{16\alpha a_0^3\mathcal{N}_1^2\mathcal{N}_2^2}}{\pi }{(\hbar {\omega _{LO}})^2}{(\frac{\hbar }{{2{m^*}{\omega _{LO}}}})^{1/2}}\nonumber\\
&&\times |\int {\int_0^{{q_c}} {dr{r^2}\frac{{\sin (qr)}}{{qr}}{{(\frac{r}{{{a^*}}})}^{{\mu _1} + {\mu _2} - 2}}} }\nonumber\\
&&\times \exp ( - \frac{r}{{{\vartheta _1}{a^*}}} - \frac{r}{{{\vartheta _2}{a^*}}})qdq{|^2},
\end{eqnarray}
for electron-phonon interaction\cite{wx moxing1,wx moxing2}. $k$ denotes the electronic wave vector and $a_0$ is the lattice constant.

According to the model proposed by Ridley\cite{wx S1,wx S2,wx S3}, HR factor between quantum defects can be calculated by
\begin{equation}
\mathcal{S}_{\lambda}=\sum\limits_q {\frac{{{{\left| \Delta_{\lambda}  \right|} ^2}}}{{2{{(\hbar {\omega _{LO}})}^2}}}},
\end{equation}
with the magnitude parameter
\begin{equation}
 \Delta_{-}    = \mathcal{M}(q)\int {[\varphi } _i^*(r){e^{iq \cdot r}}{\varphi _i}(r) - \varphi _j^*(r){e^{iq \cdot r}}{\varphi _j}(r)]{\rm{d}}{\bf{r}},
\end{equation}
and
\begin{equation}
 \Delta_{+}    = \mathcal{M}(q)\int {[\varphi } _i^*(r){e^{iq \cdot r}}{\varphi _i}(r) + \varphi _j^*(r){e^{iq \cdot r}}{\varphi _j}(r)]{\rm{d}}{\bf{r}}.
\end{equation}
Substituting Eq. (9) into Eqs. (16) and (17), then performing the integral, one can get
\begin{eqnarray}
 \Delta_{-} &=& \mathcal{M}(q)\{ \frac{{\sin (2{\mu _1}{\theta _1})}}{{2{\mu _1}(q{a^*}{\vartheta _1}/2){{[1 + {{(q{a^*}{\vartheta _1}/2)}^2}]}^{{\mu _1}}}}}\nonumber\\
&&- \frac{{\sin (2{\mu _2}{\theta _2})}}{{2{\mu _2}(q{a^*}{\vartheta _2}/2){{[1 + {{(q{a^*}{\vartheta _2}/2)}^2}]}^{{\mu _2}}}}}\},
\end{eqnarray}
and
\begin{eqnarray}
\Delta_{+} &=& \mathcal{M}(q)\{ \frac{{\sin (2{\mu _1}{\theta _1})}}{{2{\mu _1}(q{a^*}{\vartheta _1}/2){{[1 + {{(q{a^*}{\vartheta _1}/2)}^2}]}^{{\mu _1}}}}}\nonumber\\
&&+ \frac{{\sin (2{\mu _2}{\theta _2})}}{{2{\mu _2}(q{a^*}{\vartheta _2}/2){{[1 + {{(q{a^*}{\vartheta _2}/2)}^2}]}^{{\mu _2}}}}}\},
\end{eqnarray}
where ${\theta _1} = \arctan (q{a^*}{\vartheta _1}/2)$ and ${\theta _2} = \arctan (q{a^*}{\vartheta _2}/2)$. Converting the summation of $q$ into the integral, HR factors can be rewritten as
\begin{widetext}
	\begin{eqnarray}
	\mathcal{S}_{-}&=&\frac{{\alpha {q_c}}}{\pi }{[\frac{\hbar }{{2{m^*}{\omega _{LO}}}}]^{1/2}}\int_0^1 {\frac{{{{\sin }^2}[{b_1}\arctan ({c_1}x)]}}{{{{({b_1}{c_1}x)}^2}{{[1 + {{({c_1}x)}^2}]}^{{b_1}}}}}} dx+ \frac{{\alpha {q_c}}}{\pi }{[\frac{\hbar }{{2{m^*}{\omega _{LO}}}}]^{1/2}}\int_0^1 {\frac{{{{\sin }^2}[{b_2}\arctan ({c_2}x)]}}{{{{({b_2}{c_2}x)}^2}{{[1 + {{({c_2}x)}^2}]}^{{b_2}}}}}} dx\nonumber\\
	&&- \frac{{2\alpha {q_c}}}{\pi }\int_0^1 {{{[\frac{\hbar }{{2{m^*}{\omega _{LO}}}}]}^{1/2}}} \frac{{\sin [{b_1}\arctan ({c_1}x)]}}{{({b_1}{c_1}x){{[1 + {{({c_1}x)}^2}]}^{{b_1}/2}}}}\times \frac{{\sin [{b_2}\arctan ({c_2}x)]}}{{({b_2}{c_2}x){{[1 + {{({c_2}x)}^2}]}^{{b_2}/2}}}}dx.
	\end{eqnarray}
\end{widetext}
for the transitions between the same types of defects, and
\begin{widetext}
	\begin{eqnarray}
	\mathcal{S}_{+}&=&\frac{{\alpha {q_c}}}{\pi }{[\frac{\hbar }{{2{m^*}{\omega _{LO}}}}]^{1/2}}\int_0^1 {\frac{{{{\sin }^2}[{b_1}\arctan ({c_1}x)]}}{{{{({b_1}{c_1}x)}^2}{{[1 + {{({c_1}x)}^2}]}^{{b_1}}}}}} dx+ \frac{{\alpha {q_c}}}{\pi }{[\frac{\hbar }{{2{m^*}{\omega _{LO}}}}]^{1/2}}\int_0^1 {\frac{{{{\sin }^2}[{b_2}\arctan ({c_2}x)]}}{{{{({b_2}{c_2}x)}^2}{{[1 + {{({c_2}x)}^2}]}^{{b_2}}}}}} dx\nonumber\\
	&&+ \frac{{2\alpha {q_c}}}{\pi }\int_0^1 {{{[\frac{\hbar }{{2{m^*}{\omega _{LO}}}}]}^{1/2}}} \frac{{\sin [{b_1}\arctan ({c_1}x)]}}{{({b_1}{c_1}x){{[1 + {{({c_1}x)}^2}]}^{{b_1}/2}}}}\times \frac{{\sin [{b_2}\arctan ({c_2}x)]}}{{({b_2}{c_2}x){{[1 + {{({c_2}x)}^2}]}^{{b_2}/2}}}}dx.
	\end{eqnarray}
\end{widetext}
for the transitions between different types of defects. ${q_c} = \sqrt[3]{{6{\pi ^2}}}/{a_0}$ is the radius of sphere with Brillouin zone volume. ${b_1} = 2{\mu _1}$, ${b_2} = 2{\mu _2}$, ${c_1} = {q_c}{a^*}{\vartheta _1}/2$, ${c_2} = {q_c}{a^*}{\vartheta _2}/2$, and $q = q_cx$.  In this paper, we choose one of typical MHPs CH$_3$NH$_3$PbI$_3$ (MAPbI$_3$) as an example to discuss these transition processes. The values of parameters for the MAPbI$_3$ are given in Table I.

\begin{figure*}[htbp]
	\centering
	\includegraphics[width=6in,keepaspectratio]{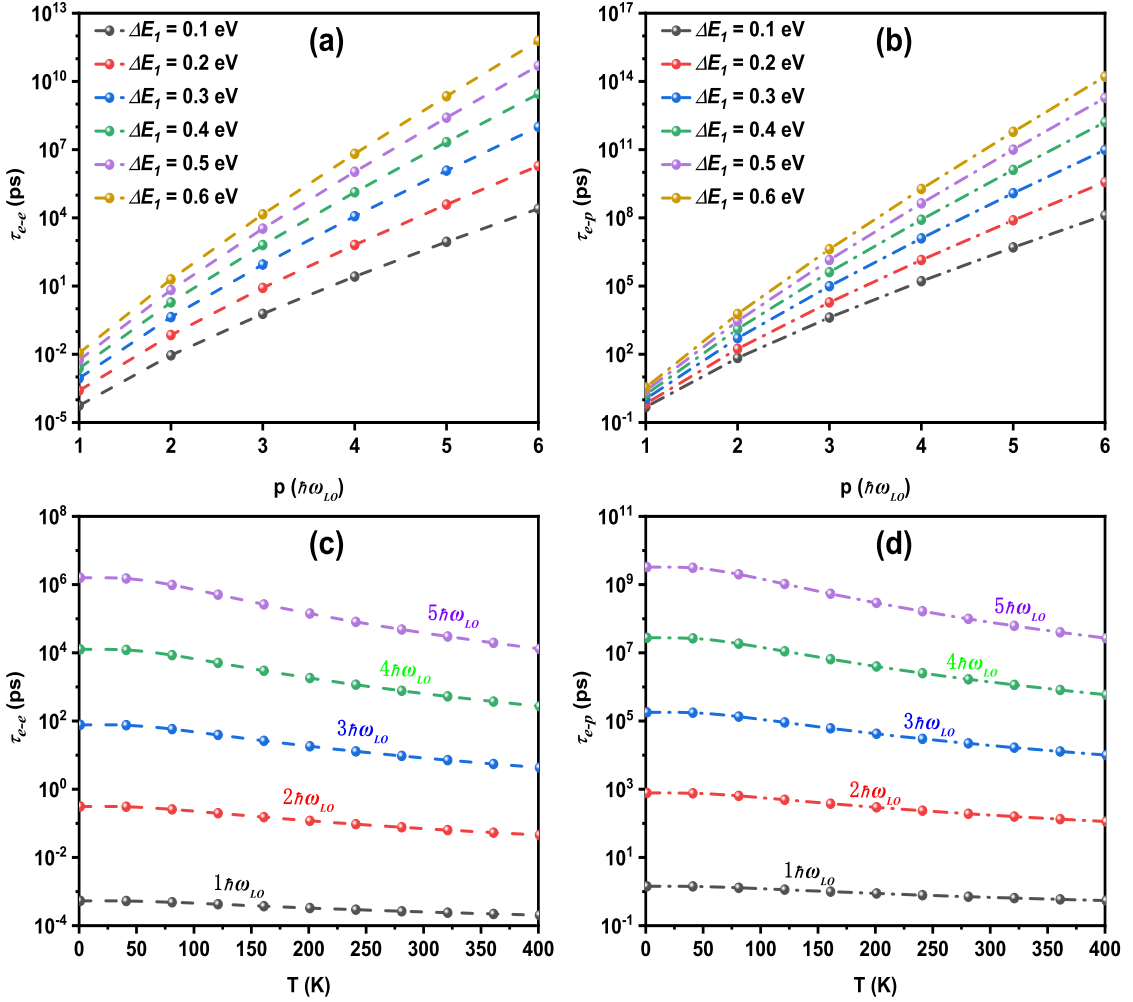}
	\caption{\label{compare} Lifetimes of the non-radiative multiphonon energy resonance transfer between donor-like defects as a function of the emitted phonon number $p$ in the mechanisms of the electron-electron interaction $\tau_{e-e}$ (a) and the electron-phonon interaction $\tau_{e-p}$ (b) for the different depths of the initial defect $\Delta E_1$ at $T$ = 300 K, where the final defect depth $\Delta E_2$ satisfies the relation of $\Delta E_2=\Delta E_1+p\hbar {\omega _{LO}}$. The temperature dependences of lifetime for different phonons emission  between donor-like defects in the mechanisms of the electron-electron interaction (c) and the electron-phonon interaction (d), in which a fixed initial defect depth $\Delta E_1 = 0.2$ eV is selected.}
\end{figure*}

\section{results and discussion}
HR factor is a key parameter determining the lifetime of ERT, which was originally proposed by Huang for F-centres in ionic crystals in 1950\cite{wx Huang1} and denotes the average phonon numbers around F-centres, reflecting the magnitude of lattice distortion arising from the strong coupling between electronic states and lattice vibration (phonons). The magnitude could be expressed by the ``phonon cloud" and its energy is described by $\mathcal{S}_{\lambda}\hbar {\omega _{LO}}$, namely the lattice relaxation energy. The dependences of HR factors ($\mathcal{S}_{-}$) on the depth difference between initial ($\Delta E_1$) and final ($\Delta E_2$) defect levels are shown in Figs. 2 (a) and (b) for the donor-donor and acceptor-acceptor transitions, respectively. One can see that $\mathcal{S}_{-}$ increases with increasing the depth separation between $\Delta E_1$ and $\Delta E_2$. However, its value only steps up to 0.25 and 1.25 for the donor-like and acceptor-like defects, respectively. Although a single quantum defect exhibits a significant lattice relaxation effect (namely, the big value of HR factor)\cite{wx moxing1}, the relative lattice relaxation effect between two defect levels in the same type is not strong enough, which is, in fact, not beneficial to the non-radiative processes between two defect levels. On the contrary, HR factors ($\mathcal{S}_{+}$) for donor-acceptor transitions can reach in the range of $26\sim32$, depending on the depth separation between donor-like and acceptor-like defect shown in Fig. 2 (c). This strong lattice relaxation effect, stemming from the sum of the contributions from the electron and hole overlapping with each other, will promote defect-assisted non-radiative transitions.

\begin{figure*}[htbp]
	\centering
	\includegraphics[width=6in,keepaspectratio]{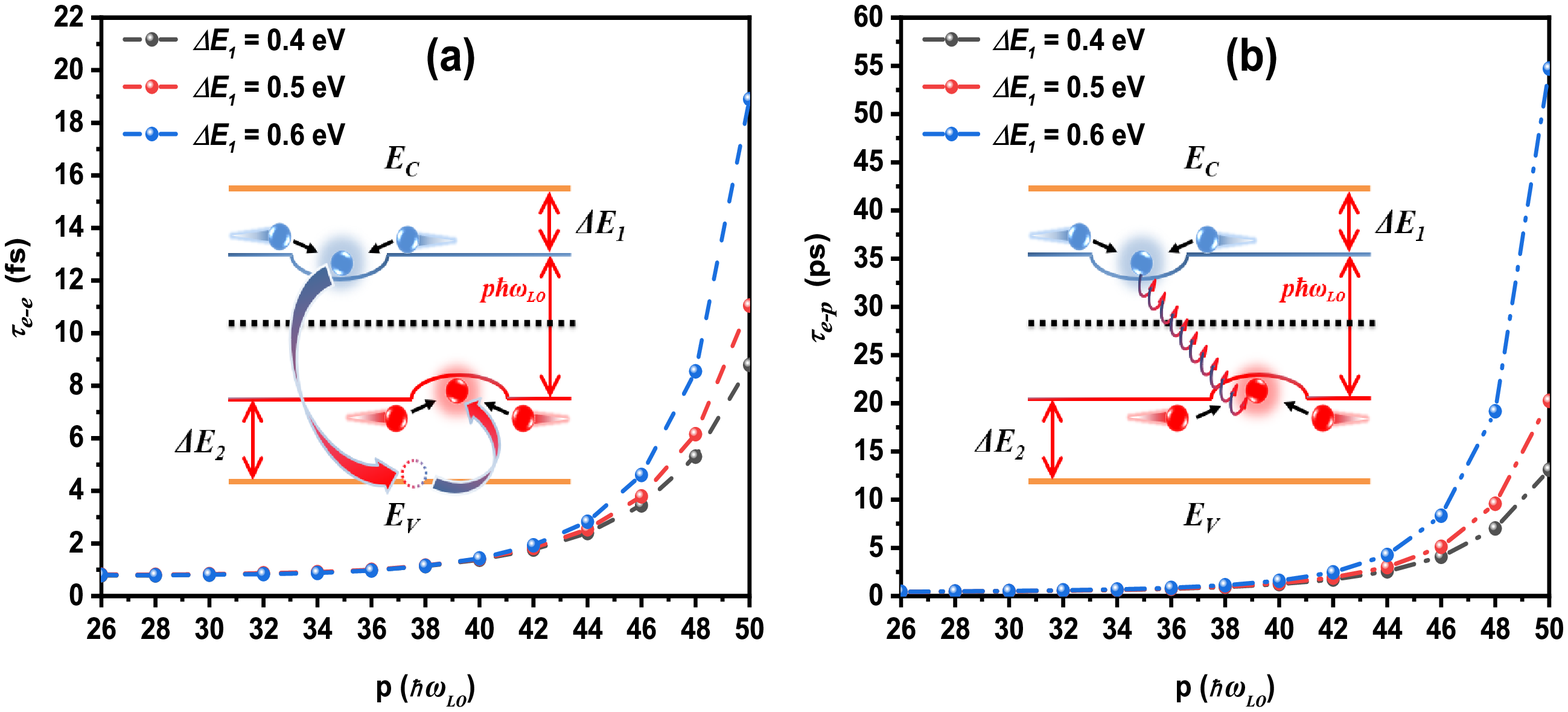}
	\caption{\label{compare} Lifetimes of the non-radiative multiphonon energy resonance transfer between donor- and acceptor-like defects as a function of the emitted phonon number $p$ in the mechanisms of the electron-electron interaction (a) and the electron-phonon interaction (b) for the different depths of the initial defect $\Delta E_1$ at $T$ = 300 K. The final defect depth $\Delta E_2$ satisfies the relation of $\Delta {E_2} = \left| {{E_C} - {E_V}} \right| - \Delta {E_1} - p\hbar {\omega _{LO}}$, where $\left| {{E_C} - {E_V}} \right|$ is the bandgap.}
\end{figure*}

We first analyze the dependences of non-radiative multiphonon ERT between donor-like defects on the emitted phonon number $p$ and the initial defect depth $\Delta E_1$, where the final defect depth satisfies the relation of $\Delta E_2=\Delta E_1+p\hbar {\omega _{LO}}$. The multiphonon ERT induced by electron-electron and electron-phonon interactions are shown in Figs. 3(a) and (b), respectively. As can be seen that the lifetime ($\tau_{e-e}$ or $\tau_{e-ph}$ ) has an obvious increase of more than one order of magnitude when $p$ increases by one, which stems from the weak lattice relaxation effect between the same types of defects as given in Fig. 2(a). With the increasing of the initial defect depth $\Delta E_1$, these processes of multiphonon ERT are also significantly suppressed. For example, the lifetime can maintain in the range of hundreds of picoseconds with the successive emitting five phonons at $\Delta E_1=0.1$ eV as shown in Fig. 3(a) induced by electron-electron interaction. But, at $\Delta E_1=0.6$ eV, the process of three phonons already reaches tens of nanoseconds, which indicates that charge carries can transfer effectively by multiphonon processes between the neighbouring shallow defect levels, causing the decrease of the carrier concentration seriously. Moreover, from the comparisons between Figs. 3(a) and (b), one can see that the electron transfer processes induced by the electron-electron interaction are much more faster (about three orders of magnitude) than the same processes induced by the electron-phonon interaction, which reveals that the electron-electron interaction will dominate the non-radiative ERT processes among quantum defects. The temperature dependence of lifetimes of these processes are plotted in Figs. 3(c) and (d) for the electron-electron and electron-phonon interactions, respectively. One can see that lifetimes of these multiphonon processes become shorter with the increasing of the temperature, which means that temperature increasing makes more phonons are excited, resulting in carriers are more likely to be captured by deeper defects. The similar results are also obtained for acceptor-like defects and neutral defects shown in the supplemental materials.

\begin{table}[t!]
	\caption{\label{compare} Huang-Rhys factor $\mathcal{S}_{+}$ are calculated by Eq. (21) for the different depths of the final acceptor-like defect $\Delta E_2$. The depth of the initial donor-like defect $\Delta E_1$=0.4 eV is fixed.}
	\centering	 \begin{tabular}{m{2cm}<{\centering}m{1cm}<{\centering}m{1cm}<{\centering}m{1cm}<{\centering}m{1cm}<{\centering}m{1cm}<{\centering}m{2cm}<{\centering}m{2cm}<{\centering}}
		 \hline
         Final defect depth ($\Delta E_2$)  & 0.7875 eV  &  0.705 eV & 0.54 eV & 0.4575 eV\\\hline
		 Phonon number ($p$)                & 25  &  30 & 40 & 45\\\hline
		 HR factor ($\mathcal{S}_{+}$)  &   31.7  &   31.2 & 30 & 29.2\\
	    \hline
	\end{tabular}
\end{table}
\begin{figure*}[htbp]
	\centering
	\includegraphics[width=7in,keepaspectratio]{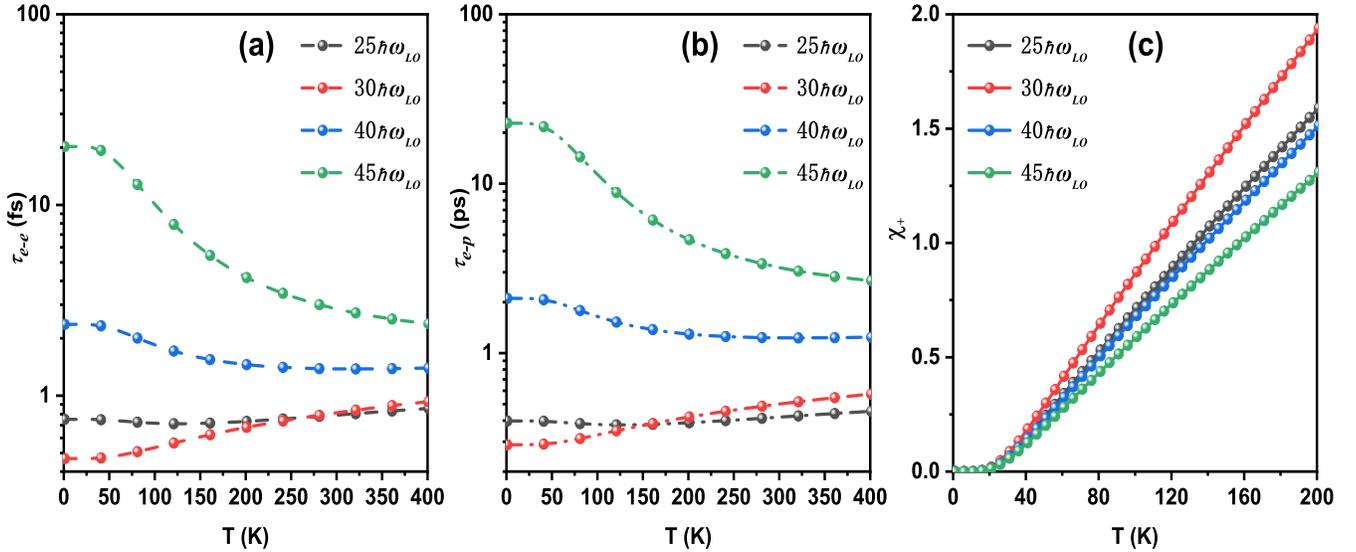}
	\caption{\label{compare} The temperature dependence of the lifetimes of multiphonon processes between donor- and acceptor-like defects for the different phonons $p$ emission in the mechanisms of the electron-electron interaction (a) and the electron-phonon interaction (b), where a fixed depth of the initial defect  $\Delta E_1 = 0.4$ eV is selected. (c) The parameter $\chi_+$ as a function of the temperature $T$ for the different phonon number $p$.}
\end{figure*}

Figs. 4(a) and (b) show the non-radiative ERT processes between donor- and acceptor-like defects for the mechanisms of electron-electron interaction and electron-phonon interaction, respectively, in which the relationship between the initial and final defect depths satisfies $\Delta {E_2} = \left| {{E_C} - {E_V}} \right| - \Delta {E_1} - p\hbar {\omega _{LO}}$, where $\left| {{E_C} - {E_V}} \right|$ is the band gap. It can be seen that these non-radiative multiphonon processes are very fast for different initial (donor-like) defect depths $\Delta E_1=$ 0.4 eV, 0.5 eV, and 0.6 eV. Even at the large emitted phonon number $p=50$, the lifetime varies just only in one order of magnitude scale, which arises from the strong lattice relaxation effect (namely the large HR factors) as shown in Fig. 2 (c). These ultrafast processes will result in a dramatic reduction of free carriers concentration in a short time, which reveal that the charge carriers can be transferred effectively via multiphonon processes between donor- and acceptor-like defects at any depth. Recently, Merdasa $et$ $al.$ observed large fluctuations of luminescence intensity, which was often referred to as a blinking phenomenon in MAPbI$_3$ by luminescence super-resolution imaging and spectroscopy\cite{wx r6}. They ascribed this phenomenon to the ultrafast non-radiative recombination via the ``supertraps" mechanism (donor-acceptor pairs). Meanwhile, the non-radiative process via these ``supertraps" at low defect concentrations was much more efficient than via other single defect states at higher concentrations. Therefore, these rapid non-radiative multiphonon transitions from donor- to acceptor-like defects obtained in the present model could provide a possible explanation for this mechanism of ``supertraps".

The temperature dependences of multiphonon transitions from donor- to acceptor-like defects are shown in Figs. 5(a) and (b) for the mechanisms of electron-electron interaction and electron-phonon interaction, respectively. One can find that the variational trends of the lifetime with temperature depend on the comparison of HR factor $\mathcal{S}_{+}$ with $p$ (the relationship between $\mathcal{S}_{+}$ and $p$ is shown in Table II). When $\mathcal{S}_{+}$ is bigger than $p$, the lifetime increases with temperature. This is because the relaxation energy ($\mathcal{S}_{+}\hbar {\omega _{LO}}$) are inclined to relax into the environment directly, but the enhancement of lattice vibration by the temperature rising disturbs the direct relaxation processes. When $\mathcal{S}_{+}$ is smaller than $p$, the relaxation processes depend on the temperature exciting more phonons to compensate the difference ($p-\mathcal{S}_{+}$), and thus results in multiphonon processes become faster with temperature. In addition, all the curves vary hardly in low temperature region ($T$$<$25 K),  which is very consistent with the temperature turning point of $\chi_{+}$ as shown in Fig. 5(c). This suggests that $\chi_{+}$ is also a key parameter to determine the temperature dependence of these non-radiative multiphonon processes in this model. Finally, we must emphasize that MAPbI$_3$ undergoes two structural phase transitions with temperature\cite{wx phase1,wx phase2,wx phase3}, which has the significant influence on the strength of the electron-phonon interaction and phonon energy. These effects are not taken into account in the present paper.

\section{conclusion}
In conclusion, we theoretically study the non-radiative multiphonon processes of energy resonance transfer among quantum defects for the induced mechanisms of the electron-electron interaction and electron-phonon interactions, respectively. We find that (1) charge carrier can transfer by non-radiation multiphonon processes between the neighboring levels of the same type shallow defects, but it will be suppressed with the increasing of the defect depth; (2) non-radiation multiphonon processes between donor- and acceptor-like defects are very fast and  independence of the defect depth, which provides a possible explanation for the ``supertraps"of the charge carriers in recent experiment; (3) the comparison of Huang-Rhys factor $\mathcal{S_+}$ with the emitted phonon number $p$ plays an important role in determining the variational trends of temperature dependence of these multiphonon processes.

This work was supported by National Natural Science Foundation of China (Grand No.  11674241).

\end{document}